\documentstyle[12pt,epsf]{article}
\newcommand{\be}{\begin{equation}}
\newcommand{\ee}{\end{equation}}
\newcommand{\bea}{\begin{eqnarray}}
\newcommand{\eea}{\end{eqnarray}}
\newcommand{\ba}{\begin{array}}
\newcommand{\ea}{\end{array}}
\newcommand{\bean}{\begin{eqnarray*}}
\newcommand{\eean}{\end{eqnarray*}}
\newcommand{\ct}{\cite}
\newcommand{\r}{\ref}
\newcommand{\n}{\nonumber}
\newcommand{\vs}{\vspace}
\def\l{\label}
\def\a{\alpha}
\def\b{\beta}
\def\c{\gamma}
\def\d{\delta}
\def\e{\epsilon}
\def\ve{\varepsilon}

\def\o{\omega}
\def\tr{{\rm Tr}}

\def\t{\tilde}
\def\wt{\widetilde}

\def\ca{{\cal A}}
\def\cb{{\cal B}}
\def\cc{{\cal C}}
\def\cd{{\cal D}}
\def\cf{{\cal F}}
\def\bm{{\bf m}}
\def\bn{{\bf N}}
\def\bl{{\bf L}}
\def\bz{{\bf Z}}
\def\2t{{\bf T}^2}
\def\3t{{\bf T}^3}
\def\4t{{\bf T}^4}
\makeatletter
\begin{document}

\topmargin 0pt

\oddsidemargin -3.5mm

\headheight 0pt

\topskip 0mm \addtolength{\baselineskip}{0.20\baselineskip}
\begin{flushright}
KIAS-P99112 \\
SOGANG-HEP 265/99 \\
UTTG-10-99 \\
{\tt hep-th/9912272}
\end{flushright}
\vs{5mm}
\begin{center}
{\large \bf Twisted Bundles on Noncommutative $\4t$  and
  D-brane Bound States}\\
\vs{5mm}
Eunsang Kim~\footnote{eskim@wavelet.hanyang.ac.kr},
Hoil Kim~\footnote{hikim@gauss.kyungpook.ac.kr},
Nakwoo Kim~\footnote{N.Kim@qmw.ac.uk},
Bum-Hoon Lee~\footnote{bhl@ccs.sogang.ac.kr},\\
Chang-Yeong Lee~\footnote{leecy@zippy.ph.utexas.edu},
and Hyun Seok Yang~\footnote{hsyang@physics4.sogang.ac.kr}\\
\vs{5mm}
${}^1${\it Department of Industrial and Applied Mathematics,
Kyungpook National University,\\Taegu 702-701, Korea}\\

${}^2${\it Topology and Geometry Research Center, Kyungpook National University,\\
Taegu 702-701, Korea}\\

${}^3${\it Physics Department, Queen Mary and Westfield College, Mile End Road,\\
London E1 4NS, UK}\\

${}^{4,6}${\it Department of Physics, Sogang University, Seoul 121-742, Korea}\\

${}^5${\it Theory Group, Department of Physics, University of Texas,
 Austin, TX 78712, USA and\\
 Department of Physics, Sejong University, Seoul 143-747, Korea}\\
\vs{5mm}
{\bf ABSTRACT}
\end{center}
We construct twisted quantum bundles and adjoint sections
 on noncommutative $\4t$,
and investigate relevant D-brane bound states with 
 non-Abelian backgrounds.
We also show that the noncommutative $\4t$ with non-Abelian backgrounds
exhibits SO$(4,4|\bz)$ duality and via this duality we get a 
 Morita equivalent $\4t$ on which only D$0$-branes exist.
For a reducible non-Abelian background, the moduli space of D-brane
bound states in Type II string theory takes the form
$\prod_a (\4t)^{q_a}/S_{q_a}$.
\vs{5mm}
\begin{flushleft}
December 1999 \\
\end{flushleft}

\newpage
\section{Introduction}
Recent developments in nonperturbative string theories have
provided new powerful tools to understand supersymmetric gauge
theories \ct{gk}. The Bogomol'nyi-Prasad-Sommerfeld (BPS) brane
configurations led to many exact results on the vacuum
structure of supersymmetric gauge theories.
One may be interested in counting degeneracy of D-brane bound states
of type II string theory compactified on ${\bf R}^{1,9-d} \times X$ in
which a gauge field strength $F$ and a Neveu-Schwarz B field on the
brane are nonzero. Then $p$-branes wrapped on a compact $p$-cycle $W_p
\subset X$ and their bound states look like particles in the effective 
${\bf R}^{1,9-d}$ spacetime. Moreover, the degeneracy of the bound
states is the same as the number of ground states in the corresponding
quantum field theory on the D-brane worldvolume \ct{witt95}.

The D-brane moduli space \ct{dm,hm} can be defined as a space of Chan-Paton
vector bundle $E$ over $X$ or a space of solutions to the equation
given by 
\[\d \lambda= F_{MN}\Gamma^{MN}\xi+ \eta=0\]
for some pair of covariantly constant spinors $\xi$ and $\eta$ on 
${\bf R}^{1,9-d} \times X$.
The various RR charges are given by the Mukai vector $Q=v(E)={\rm{Ch}}(E)
\sqrt{\hat A(X)}\in H^{2*}(X,{\bf Z})$ where 
${\rm Ch}(E)={\rm{Tr\;exp}}\left[\frac{1}{2\pi}(F-B)\right]$ is the Chern character
and ${\hat A(X)}=1-p_1(X)/24$ is the A-roof genus for four dimensional
manifold $X$.
Then the supersymmetric, BPS bound states, for example (D0, D2, D4)
bound states on $\4t$ or ${\bf K3}$, are allowed by the
Chern-Simons couplings \ct{ghm}
\[ \int_{X \times {\bf R}} C^{RR}\wedge Q.\]

It was shown in \ct{cds,dh} that noncommutative geometry can be
successfully applied to the compactification of M(atrix) theory
\ct{bfss} in a certain background. In those papers, it was argued that 
M(atrix) theory in a 3-form potential background with
one index along the lightlike circle and 2 indices along ${\bf T}^d$ 
is a gauge theory on noncommutative torus, specifically
(d+1)-dimensional noncommutative super Yang-Mills (NCSYM) theory.
Many more discussions of M- and string theory compactifications on
these geometries followed, for example 
\ct{ncg,riesch,sch,ho,brmozu,hove,sw}.

One obvious advantage of NCSYM
theory defined on ${\bf T}^d$ is that the T-duality, SO$(d,d|\bz)$,
of type II string theory compactified on torus 
becomes manifest \ct{cds,dh,brmozu,hove,sw}.
Morita equivalence between two noncommutative torus \cite{riesch,sch} 
encompasses the Nahm transformation part of T-duality,
not clearly observed in conventional Yang-Mills theory.
Using this symmetry, it may be possible to systematically count D-brane
bound states on $\4t$ or ${\bf K3}$ as ground state configurations for
the supersymmetric gauge theory. 

For compactifications on $\2t$ and $\3t$, 
generic $U(N)$ bundles on it admit vanishing SU$(N)$ curvature
\ct{cds,ho,brmozu}.
However, for compactifications on tori of
dimension 4 or larger,
not all bundles allow vanishing $SU(N)$ curvature 
so we have to consider more generic bundles with 
nonvanishing $SU(N)$ curvature. 
It turns out \ct{tH79,hooft} that one can construct twisted 
$SU(N)$ gauge bundle on $\4t$ with fractional instanton number.
However, in discussing the $U(N)$ gauge theory as D-brane dynamics,
it is understood that the total instanton number is integral since
the instanton number is related to D0-brane charges inside D4-branes,
which should satisfy Dirac quantization due to the existence of
D6-brane in type IIA string theory \ct{polchinski}. 
In \ct{guram,d-brane}, 't Hooft solutions on twisted bundles on commutative tori
were realized by D-brane configurations (D-brane bound states)
wrapped on tori in type II string theory, and it was shown that
U-duality relates their bound states.

In general one can consider gauge bundle on $\4t$ with non-Abelian 
constant curvature \ct{hooft}.
In that case, non-Abelian backgrounds can be 
obviously supersymmetric for self-dual or anti-self-dual fields
since the supersymmetry of $D$-brane world volume theory
may be given by
\[\d \lambda= F_{MN}\Gamma^{MN}\xi.\]  
Thus, in order to study the BPS spectrums of the NCSYM theory 
on the non-Abelian backgrounds, 
it will be useful to construct the corresponding gauge bundles.
In the presence of non-Abelian backgrounds as well as Abelian backgrounds,
the gauge bundle may be twisted by the background magnetic fluxes. 
While Abelian backgrounds universally twist $U(N)$ gauge bundle,
in the case of non-Abelian backgrounds where
the magnetic fluxes in $U(N)$ are decomposed into $U(k)$ part 
and $U(l)$ part \ct{hooft},
the magnetic flux in $U(k)$ part twists $U(k) \subset U(N)$ gauge bundle and 
that in $U(l)$ part does $U(l) \subset U(N)$ gauge bundle.
This causes two different deformation parameters to appear.

The Chern character maps K-theory to cohomology 
i.e. $\rm{Ch }:K^0(X)\rightarrow H^{\rm{even}}(X,\bz)$ and 
$K^1$ to odd cohomology and $\rm{Ch}(E)=
\rm{Ch}_0(E)+\rm{Ch}_1(E)+\rm{Ch}_2(E)$ when $X$ is
4-dimensional and $E$ is a vector bundle over $X$. 
Here $\rm{Ch}_0(E)$ is the rank of $E$, $\rm{Ch}_1(E)$ is the first
Chern class and $\rm{Ch}_2(E)$ corresponds to the instanton number. 
$\rm{Ch}_1(E)$ is integral winding number when the torus is
commutative and it is not integer anymore when the torus is
noncommutative but $\rm{Ch}_2(E)$ still remains integral 
even if the torus becomes noncommutative \ct{riesch,sch,hove}.
However D-brane charges take values in $K(X)$, the K-theory of $X$
\ct{K}, which constitutes a group of integer $\bz$.
The (4+1)-dimensional $U(N)$ SYM theory 
can be interpreted as dynamics of $N$ D4-branes. Six magnetic fluxes
are D2-branes wound around 6 two-cycles of $\4t$. Instantons
are D0-branes bound to D4-branes. 
Thus, even when NS-NS two-form potential background
is turned on, the physical D-brane numbers should be integers.
In addition, the rank, 6 fluxes, and instanton (altogether, 
eight components) make 
a fundamental multiplet of the Weyl spinor representation of
$SO(4,4|\bz)$ \ct{hove}.

Since the explicit constructions of twisted bundles and adjoint
sections in the literatures have been performed only for Abelian
backgrounds, we will construct them for constant 
non-Abelian backgrounds in this paper. 
In section 2, we construct twisted bundles
on noncommutative $\4t$. In section 3, adjoint sections on the twisted
bundle will be constructed. In section 4, we show that the modules of 
D-brane bound states exhibit an $SO(4,4|\bz)$ duality 
and the action of this group gives 
Morita equivalent $\4t$ on which only $D0$-branes exist.
Section 5 devotes conclusion and comments on our results.
In appendix, we present some details of the representation of
$SO(4,4|\bz)$ Clifford algebra.

\section{Twisted Quantum Bundles On $\4t$}

To define the noncommutative geometry, we understand 
the space is noncommutative, viz.
\be
\l{ncst}
[x^\mu , x^\nu ] = -2\pi i \Theta_{\mu\nu}.
\ee
Then the noncommutative $\4t$, which will be denoted by $\4t_{\Theta}$, is
generated by translation operators $U_\mu$ defined by $U_\mu=e^{ix_\mu}$
and satisfies the commutation relation 
\be
\l{endom}
U_\mu U_\nu =e^{2\pi i\Theta_{\mu\nu}}
U_\nu U_\mu.
\ee
Also, we introduce partial derivatives satisfying
\[[\partial_\mu,x^\nu]=\d_\mu^\nu, \quad
[\partial_\mu,\partial_\nu]=0.\]

We construct quantum $U(N)$ bundles on $\4t_{\Theta}$ 
following the construction of \cite{ho,brmozu} and \ct{hooft}. 
Start with a constant curvature connection
\be
\l{connect}
\nabla_\mu = \partial_\mu + i F_{\mu\nu} x^\nu,
\ee
where Greek indices run over spatial components only. 
In this paper we allow the $U(N)$ gauge fields with 
nonvanishing $SU(N)$ curvature in order to consider
non-Abelian backgrounds. 
Following the ansatz taken by 't Hooft \ct{hooft}, we take the
curvature $F_{\mu\nu}$ as the Cartan subalgebra element:
\be
\l{bF}
F_{\mu\nu}= F_{\mu\nu}^{(1)} + F_{\mu\nu}^{(2)}, 
\ee
where $F_{\mu\nu}^{(1)}=\tr F_{\mu\nu}$ and  $F_{\mu\nu}^{(2)} \in
u(1) \subset su(N)$.
The constant curvature is given by
\be
{\cal F}_{\mu\nu} = i [ \nabla_\mu , \nabla_\nu ].
\ee
And one can calculate to get
\be
\l{F}
{\cal F} = ( 2F + 2\pi F \Theta F).
\ee
Note that both $F$ and $\Theta$ are antisymmetric $4\times 4$ matrices.

The gauge transformations of fields in the adjoint representation of
gauge group are insensitive to the center of the group, 
e.g. ${\bf Z}_N$ for $SU(N)$. Thus, for the adjoint fields in $SU(N)$
gauge theory, it is sufficient to consider the gauge group as being 
$SU(N)/{\bf Z}_N$. However, there can be an obstruction to go from an
$SU(N)/{\bf Z}_N$ principal fiber bundle to an $SU(N)$ bundle if the
second homology group of base manifold $X$, $H_2(X,{\bf Z}_N)$, does
not vanish \ct{sed}.   
In order to describe such a nontrivial $U(N)$ bundle, it is helpful to
decompose the gauge group into its Abelian and non-Abelian components
\be
\l{un} 
U(N)=\Bigl( U(1) \times SU(N) \Bigl)/\bz_N.
\ee
It means that we identify an element $(g_1, g_N) \in U(1) \times
SU(N)$ with $(g_1c^{-1},c g_N)$, where $c \in \bz_N$.
Therefore one can arrange the twists in $U(N)$ to be trivial
by cancelling them between $SU(N)$ and $U(1)$ \ct{guram}.
This requires consistently combining solutions of $SU(N)/\bz_N$ with
$U(1)$ solutions as to cancel the total twist.

To characterize the generic $U(N)$ gauge bundle on $\4t_{\Theta}$, 
we allow the gauge bundle be periodic up to gauge transformation
$\Omega_\mu$, i.e.
\be
\nabla_\mu (x^\alpha + 2\pi \delta^\alpha_\nu )
= \Omega_\nu (x^\alpha) \nabla_\mu (x^\alpha) \Omega_\nu^{-1} (x^\alpha).
\label{periodic}
\ee
Consistency of the transition functions of the $U(N)$ bundle requires
the so-called cocycle condition
\be
\l{cocycle}
\Omega_\mu (x^\alpha + 2\pi \delta^\alpha_\nu) \Omega_\nu (x^\alpha) =
\Omega_\nu (x^\alpha + 2\pi \delta^\alpha_\mu) \Omega_\mu (x^\alpha).
\ee
However the $SU(N)$ transition function $\wt{\Omega}_\mu (x^\alpha)$
may be twisted as \ct{tH79}
\be
\l{tcocycle}
\wt{\Omega}_\mu (x^\alpha + 2\pi \delta^\alpha_\nu) 
\wt{\Omega}_\nu (x^\alpha) =Z_{\mu\nu}\,
\wt{\Omega}_\nu (x^\alpha + 2\pi \delta^\alpha_\mu) 
\wt{\Omega}_\mu (x^\alpha),
\ee
where $ Z_{\mu\nu}=e^{-2\pi i n_{\mu\nu}/N}$ is the center of $SU(N)$. 

Write $\Omega_\mu (x)$ as a product of an $x$-dependent part
and a constant part
\be
\l{Omega}
\Omega_\mu (x)= e^{i (P_{\mu\nu}^{(1)} +P_{\mu\nu}^{(2)}) x^\nu}W_\mu,
\ee
where $P^{(1)}_{\mu\nu}$ is antisymmetric and proportional to the
identity in the Lie algebra of $U(N)$ while
$P^{(2)}_{\mu\nu}$ is an element of $u(1) \subset su(N)$.
And constant $N\times N$ unitary matrices $W_\mu$ are taken as
$SU(N)$ solutions generated by 't Hooft clock and shift matrices. 
For comparision, our $P^{(2)}_{\mu\nu}$ in (\r{Omega}) corresponds
to the constant $SU(N)$ field strength $\alpha_{\mu\nu}$ in 't Hooft ansatz 
in \ct{hooft} if we consider commutative $\4t$.

In the case of vanishing $su(N)$ cuvature,
$F_{\mu\nu}^{(2)}=P_{\mu\nu}^{(2)}=0$,
an explicit construction of gauge bundles with magnetic and electric
fluxes was given in \ct{hove}.
 For the nonvanishing $su(N)$ curvature case,
following 't Hooft solution \ct{hooft}  
we consider diagonal connections which
break $U(N)$ to $U(k) \times U(l)$  
where each block has vanishing
$SU(k)$ and $SU(l)$ curvature. 
We also consider the groups $U(k)$
and $U(l)$ as $U(k)=(U(1) \times SU(k))/\bz_k$ and 
$U(l)=( U(1) \times SU(l))/\bz_l$, respectively. Thus the twists of
$SU(k)$ or $SU(l)$ part can be trivialized by each $U(1)$ part. 
Since the $U(1)$ in $U(N)$ is the direct sum of $U(1)$ in $U(k)$ 
and $U(1)$ in $U(l)$, the $SU(N)$ twist tensor should be a sum of 
$SU(k)$ and $SU(l)$ twist tensors.   

Here we take the generator $\sigma$ in
$u(1)\subset su(N)$ as
\be
\l{sigma}
\sigma = \pmatrix{ 
l\,{\bf 1}_k & 0\cr
0 & -k\,{\bf 1}_l},
\ee
where $k\times k$ matrix ${\bf 1}_k$ is the identity in $U(k)$ and
$l\times l$ matrix ${\bf 1}_l$ is that in $U(l)$.
Then we take the $SU(N)$ connection to be proportional to $\sigma$.
Since the $U(N)$ gauge field in (\r{connect}) contain only the matrix
$\sigma$ and the identity matrix ${\bf 1}_N$ in $U(N)$ 
and so commutes with $W_\mu$,
in checking (\ref{periodic}), $W_\mu$ are irrelevant in our situation
and we have
\be
\l{P}
P = 2\pi F ({\bf 1}_N + 2\pi \Theta F)^{-1}=
2\pi({\bf 1}_N + 2\pi F \Theta)^{-1}F,
\ee
where $P_{\mu\nu}=P_{\mu\nu}^{(1)} +P_{\mu\nu}^{(2)}$.
 From the ansatz of $\Omega_\mu$  (\ref{Omega}) and the cocycle
condition (\r{cocycle}), we obtain the following commutation relation 
for $W_\mu$  
\be
W_\mu W_\nu = e^{-2\pi i M_{\mu\nu} /N } W_\nu W_\mu,
\label{twisteat}
\ee
where $M$ is given by
\be
\l{MN}
M =M^{(1)}+M^{(2)}= N (2P - P\Theta P).
\ee
Here, an integral matrix $M^{(1)}_{\mu\nu}$ is coming from
 the trace part of $U(N)$, 
and $M^{(2)}_{\mu\nu}$ which is also integral is 
proportional to $\sigma$.

We now construct the solutions {\it a la} 't Hooft 
for bundles with a constant 
curvature background (\r{bF}) on $\4t_{\Theta}$.
The greatest common divisor of $(M_{\mu\nu},N)$ is invariant under
$SL(4,\bz)$ and we take it as $q$. Also, we assume the twist matrix
$M$ and the flux $P$ have the form of $q$ copies of 
$U(n)$ matrices $\bm$ and ${\wt P}$ defined by
\be
\l{M}
\bm= n(2{\wt P} - {\wt P}\Theta {\wt P}),
\qquad P={\bf 1}_q \otimes {\wt P},
\ee
where ${\bf 1}_q$ is a $q$-dimensional identity matrix.
In other words,
\be
\l{nm}
N=q\,n,\quad M=q \,{\bf 1}_q \otimes \bm,
\ee
where $n$ is the reduced rank. 
In this case, it is convenient to consider
transition functions $\Omega_\mu$ and $W_\mu$ as the following
block diagonal form \ct{brmozu}
\be
\l{tblock}
\Omega_\mu={\bf 1}_q\otimes \omega_\mu,\qquad W_\mu={\bf 1}_q \otimes
{\wt W}_\mu,
\ee
where $\omega_\mu$ and ${\wt W}_\mu$ belong to
 $U(n)$ and  $SU(n)$, respectively. Thus we will consider 
only one copy described by $U(n)$ transition functions $\omega_\mu$.

Let us define $SU(n)$ matrices $U$ and $V$ as follows
\be
U_{kl}= e^{2\pi i(k-1)/n}\,\delta_{k,l},\qquad
V_{kl}= \delta_{k+1,l}, \quad k,\,l=1,\cdots, n,
\ee
so that they satisfy $UV=e^{-2\pi i/n}VU$.
For $\4t_{\Theta}$ with vanishing $SU(n)$ curvature 
where we can put $F_{\mu\nu}^{(2)}=P_{\mu\nu}^{(2)}=0$,
there are solutions of the form
\be
\l{wt2}
{\wt W}_\mu= U^{a_\mu}V^{b_\mu},
\ee
where $a_\mu$ and $b_\mu$ are integers.
In order for the $U(n)$ twists to be trivial as in (\r{cocycle}), 
the $SU(n)$ twists $n_{\mu\nu}$ should be balanced with the $U(1)$ fluxes
${\bm}_{\mu\nu}=m_{\mu\nu}{\bf 1}_n$.
Thus, the equation (\ref{twisteat}) gives 
\be
\l{0inst}
n_{\mu\nu}=m_{\mu\nu}= a_\mu b_\nu - a_\nu b_\mu
\quad {\rm mod} \quad n.
\ee

In the case of commutative $\4t$, 't Hooft solutions with nonvanishing $SU(n)$
curvature are described by breaking $U(n)$ to $U(k)\times U(l)$
so that background gauge fields live along the diagonals
of the $U(k)$ and $U(l)$  \ct{hooft}. 
Here we have taken $n$ as $n=k+l$.
For $\4t_{\Theta}$, we now adopt a 't Hooft type solution given by
\be
\l{wtw}
{\wt W}_\mu= U_1^{a_\mu}V_1^{b_\mu} U_2^{c_\mu}V_2^{d_\mu},
\ee
where $a_\mu,\,b_\mu,\,c_\mu$ and $d_\mu$ are integers to be
determined. The matrices $U_{1,2}$ and $V_{1,2}$ acting in the two
subgroup $SU(k)$ and $SU(l)$ satisfy the following commutation rules
\bea
\l{tweat12}
&& U_1V_1=e^{-2\pi i /k \bf{I}_k}V_1U_1,\n \\
&& U_2V_2=e^{-2\pi i /l \bf{I}_l}V_2U_2,\\
&& [U_1,U_2]=[U_1,V_2]=[V_1,U_2]=[V_1,V_2]=0,\n
\eea
where $n\times n$ matrices ${\bf I}_k$ and
${\bf I}_l$ have the forms respectively
\be
\l{idmatrix}
{\bf I}_k = \pmatrix{ 
{\bf 1}_k & 0\cr
0 & 0 },\qquad 
{\bf I}_l = \pmatrix{ 
0 & 0\cr
0 & {\bf 1}_l}.
\ee
As discussed above, the triviality of the $U(n)$ twists requires a
balance between the $SU(n)$ twists $n_{\mu\nu}$
and the $U(1)$ fluxes ${\bm}^{(1)}_{\mu\nu}$, which leads to the 
identification $n_{\mu\nu}{\bf 1}_n={\bm}^{(1)}_{\mu\nu}$.
Similarly, since each block has vanishing $SU(k)$ or $SU(l)$ curvature,
the fluxes $m^{(k)}_{\mu\nu}$ in $U(k)$ and $m^{(l)}_{\mu\nu}$ in
$U(l)$ have to cancel the twists $n^{(k)}_{\mu\nu}$ in $SU(k)$ and 
$n^{(l)}_{\mu\nu}$ in $SU(l)$ respectively, which leads us
the identification as in (\r{0inst})
\be
\l{n=m}
n^{(k)}_{\mu\nu}=m^{(k)}_{\mu\nu},\quad
n^{(l)}_{\mu\nu}=m^{(l)}_{\mu\nu}.
\ee
Following the identification (\r{n=m}),
one can solve the total $SU(n)$ twists $n_{\mu\nu}$ in terms of two
sets of twists $n^{(k)}_{\mu\nu}$ and $n^{(l)}_{\mu\nu}$,
and the $SU(n)$ fluxes ${\bm}^{(2)}_{\mu\nu}$ as in \ct{hooft}.
Using (\r{tweat12}) the equation (\ref{twisteat}) gives
\be
\l{27}
\frac{n_{\mu\nu}}{n}{\bf 1}_n=\frac{n^{(k)}_{\mu\nu}}{k}{\bf I}_k + 
\frac{n^{(l)}_{\mu\nu}}{l}{\bf I}_l- \frac{{\bm}^{(2)}_{\mu\nu}}{n}. 
\ee
Taking the trace on the above equation, we get
\be
n_{\mu\nu}=n^{(k)}_{\mu\nu}+ n^{(l)}_{\mu\nu},
\ee
where 
\bea
\l{2m}
&&n^{(k)}_{\mu\nu}= a_\mu b_\nu - a_\nu b_\mu \quad {\rm mod}
\quad k,\n \\
&&n^{(l)}_{\mu\nu}= c_\mu d_\nu - c_\nu d_\mu \quad {\rm mod}
\quad l.
\eea
Recall that the Pfaffians given by twists $n^{(k)}_{\mu\nu}$ and
$n^{(l)}_{\mu\nu}$ satisfy 
\be
\frac{1}{8} \epsilon^{\mu\nu\alpha\beta} n_{\mu\nu}^{(k)} 
n_{\alpha\beta}^{(k)}= 0  \quad {\rm mod}
\quad k, \quad  \ \   \frac{1}{8} 
\epsilon^{\mu\nu\alpha\beta} n_{\mu\nu}^{(l)} 
n_{\alpha\beta}^{(l)} 
= 0  \quad {\rm mod} 
\quad l 
\ee
due to triviality of  the $SU(k)$ and $SU(l)$ parts.
However, the total $SU(n)$ twists may satisfy
\be
\frac{1}{8} \epsilon^{\mu\nu\alpha\beta} n_{\mu\nu}
n_{\alpha\beta} \ne 0  \quad {\rm mod}
\quad n
\ee
since it is not trivial in this construction.
And the  0-brane charge is given by
\be
\l{C}
C=k\cdot Pf(n^{(k)}/k)+l\cdot Pf(n^{(l)}/l)=C^{(k)}+C^{(l)}
\ee
which is an integer, due to the triviality
of each sector \cite{guram}.
Therefore, our construction
corresponds to $D$-brane bound states
involved with (4,\,2,\,2) or (4,\,2,\,2,\,0) system
depending on the value of $C$
in the language of \cite{guram}.
The (4,\,2,\,2) system is a bound state of 4-branes and 2-branes
with non-zero intersection number but no zero branes.
The (4,\,2,\,2,\,0) system is a bound state of 4, 2, and 0-branes 
with non-zero 2-brane intersection number.

For an explicit construction of these systems, we may choose
\[ n_{34}^{(k)}=n_{12}^{(l)}=0, \quad 
 n_{12}^{(k)} \ne 0, \quad   n_{34}^{(l)} \ne 0  \]
for (4,\,2,\,2), and
\[  n_{12}^{(k)} = p^{(k)}, \quad   n_{34}^{(k)} = k, \quad
  n_{12}^{(l)} = l, \quad   n_{34}^{(l)} = p^{(l)}   \]
for (4,\,2,\,2,\,0).
Here, the 0-brane charge in the (4220) case is 
given by $p^{(k)} +  p^{(l)}.$
Notice that in this construction, the (4,\,2,\,2) system can be contained 
in the (4,\,2,\,2,\,0) system as a special case.

Since some work in this direction in the vanishing SU(N) curvature 
case \ct{hove} was already done
via van Baal construction \ct{baal},
below we also show how we can construct a (4,\,2,\,2,\,0) system
{\it a la} van Baal in our case.

The equation (\ref{twisteat}) is covariant under $SL(4,\bz)$. Using
this symmetry we can always make the matrix $m= m^{(k)} + m^{(l)}$ 
to a standard symplectic form 
by performing a $SL(4,\bz)$ transformation 
$R$,
\be
\l{Msl4}
m=R m_0 R^T,
\ee
where we choose $m_0$ as
\be
m_0= \pmatrix{ 
0 & m_1+ m_3 & 0 & 0 \cr
-m_1- m_3 & 0 & 0 & 0 \cr
0 & 0 & 0 & m_2 \cr
0 & 0 & -m_2 & 0 }.
\ee
Since $m_0= m_0^{(k)} + m_0^{(l)}$, 
we take the matrices  $m_0^{(k)}$ and
$m_0^{(l)}$ as
\bea
\l{Mx}
\ba{ll}
m_0^{(k)}= \pmatrix{ 
0 & m_1 & 0 & 0 \cr
-m_1 & 0 & 0 & 0 \cr
0 & 0 & 0 &  m_2 \cr
0 & 0 & - m_2 & 0 },\quad
& m_0^{(l)}= \pmatrix{ 
0 & m_3 & 0 & 0 \cr
-m_3 & 0 & 0 & 0 \cr
0 & 0 & 0 & 0 \cr
0 & 0 & 0 & 0 }. 
\ea
\eea
Here we have taken a simple $U(l)$ solution for convenience.

Since we consider a special diagonal connection which
breaks $U(n)$ to $U(k) \times U(l)$ and each block has vanishing
$SU(k)$ or $SU(l)$ curvature, the twisted bundle 
can be decomposed into $U(k)$ part and $U(l)$ part and 
the construction in \ct{baal} can be applied to each part separately.
Introduce $q_i={\rm gcd}(m_i,\,k),\;l_0={\rm gcd}(m_3,\,l) \;(i=1,2)$ 
and
$k_i=k/q_i,\; l_1=l/l_0$. 
In \ct{baal}, it was shown that twist-eating solutions of the type 
\be
\label{twisteat2}
{\wt W}_\mu {\wt W}_\nu = e^{-2\pi i {\bf m}_0^{\mu\nu}/n}
{\wt W}_\nu {\wt W}_\mu,
\ee
where $\frac{{\bf m}_0^{\mu\nu}}{n}= \frac{m^{(k)}_{0\,\mu\nu}}{k}{\bf I}_k + 
\frac{m^{(l)}_{0\,\mu\nu}}{l}{\bf I}_l$,
can only exist if $k_1k_2|k$. We thus write $k=k_1k_2 k_0$. 
When this
restriction is satisfied,
it is straightforward to check that the following solution
satisfies (\ref{twisteat2}),
\bea
\l{twisteatsol}
{\wt W}_1 &=& U_{k_1}^{m_1/q_1} \otimes {\bf 1}_{k_2} \otimes {\bf 1}_{k_0}
\oplus U_{l_1}^{m_3/l_0} \otimes {\bf 1}_{l_0}  \nonumber \\
{\wt W}_2 &=& V_{k_1} \otimes {\bf 1}_{k_2} \otimes {\bf 1}_{k_0}
\oplus V_{l_1} \otimes {\bf 1}_{l_0} \nonumber \\
{\wt W}_3 &=& {\bf 1}_{k_1} \otimes U_{k_2}^{m_2/q_2} \otimes {\bf 1}_{k_0}
\oplus {\bf 1}_{l} \nonumber \\
{\wt W}_4 &=& {\bf 1}_{k_1} \otimes V_{k_2} \otimes {\bf 1}_{k_0}
\oplus {\bf 1}_{l}, 
\eea
where $SU(k_i)$ matrices $U_{k_i}$ and $V_{k_i}$ are defined as 
\bea
&&(U_{k_i})_{ab}= e^{2\pi i(a-1)/k_i}\,\delta_{a,b},\qquad
(V_{k_i})_{ab}= \delta_{a+1,b}, \quad a,\,b=1,\cdots, k_i,\n\\
&&(U_{l_1})_{cd}= e^{2\pi i(c-1)/l_1}\,\delta_{c,d},\qquad
(V_{l_1})_{cd}= \delta_{c+1,d}, \quad c,\,d=1,\cdots, l_1,
\eea
so that they satisfy $U_{k_i}V_{k_i}=e^{-2\pi i/k_i}V_{k_i}U_{k_i}$
and $U_{l_1}V_{l_1}=e^{-2\pi i/l_1}V_{l_1}U_{l_1}$.

\section{Adjoint Sections On Twisted Bundles}
According to the correspondence between a compact space $X$ and
the $C^*$-algebra $C(X)$ of continuous functions on $X$,
the entire topological structure of $X$ is encoded in the
algebraic structure of $C(X)$. Continuous sections of a vector bundle 
over $X$ can be identified with projective modules over the algebra
$C(X)$. Thus, in order to find the topological structure of the twisted 
bundle constructed in the previous section, it is necessary to
construct the sections of the bundle on $\4t_{\Theta}$. 
Furthermore as noted in \cite{cds}, 
if $D_\mu$ and $D_\mu'$ are two connections 
then the difference $D_\mu-D_\mu'$ belongs to the algebra of
endomorphisms of the $\4t_{\Theta}$-module. 
Thus an arbitrary connection $D_\mu$ can be written 
as a sum of a constant curvature connection $\nabla_\mu$, 
and an element of the endomorphism algebra:
\[D_\mu =\nabla_\mu +A_\mu.\]
From the relation (\r{periodic}), we see that $A$ is also an adjoint
section. Thus the algebra of adjoint sections can be regarded as 
the moduli space of constant curvature connections.

In this section we will analyze the structure of the adjoint sections
on the twisted bundles on $\4t$, closely following the method taken
by Brace {\it et al.} \ct{brmozu} and Hoffman and Verlinde \ct{hove}.
According to the decomposition (\r{nm}),
we take the adjoint sections of $U(N)$ as the form
\be 
\Phi(x^\mu)={\bf 1}_q \otimes {\wt \Phi}(x^\mu).
\ee
The sections ${\wt \Phi}$ on the twisted bundle of the adjoint representation
of $U(n)$ are $n$-dimensional matrices of
functions on $\4t_{\Theta}$ which is generated by (\r{endom}),
 endomorphisms of the module,
and satisfy the twisted boundary conditions
\be
\l{tbcas} 
{\wt \Phi}(x^\mu +2\pi\d^\mu_\nu)=\o_\nu{\wt \Phi}(x^\mu)\o_\nu^{-1}.
\ee

Suppose that the general solution for the $n$-dimensional matrices
${\wt \Phi}(x^\mu)$ has the following expansion 
\be 
\l{asection}
{\wt \Phi}(x^\mu)=\sum_{n_1\cdots n_4 \in \bz}
{\wt \Phi}_{n_1\cdots n_4}Z_1^{n_1}Z_2^{n_2}Z_3^{n_3}Z_4^{n_4}.
\ee 
We also try to find the solutions of the following form
\be
\l{Z}
Z_\mu =e^{i x_\nu {X^\nu}_\mu/n} \prod_{\alpha=1}^6
\Gamma_\alpha^{s_\alpha^\mu}
\ee
where $s_\alpha^\mu \;(\alpha=1,\cdots,6)$ are integers and
$X$ is a matrix to be determined. Here, according to the basis taken
in Eq. (\r{twisteatsol}), we define the $SU(n)$ matrices $\Gamma_\alpha$
as follows
\bea
\l{UV}
&&\Gamma_1 = U_{k_1} \otimes {\bf 1}_{k_2} \otimes {\bf 1}_{k_0}
\oplus {\bf 1}_{l},  \nonumber \\
&&\Gamma_2 = V_{k_1} \otimes {\bf 1}_{k_2} \otimes {\bf 1}_{k_0}
\oplus {\bf 1}_{l}, \nonumber \\
&&\Gamma_3 = {\bf 1}_{k_1} \otimes U_{k_2} \otimes {\bf 1}_{k_0}
\oplus {\bf 1}_{l}, \nonumber \\
&&\Gamma_4 = {\bf 1}_{k_1} \otimes V_{k_2} \otimes {\bf 1}_{k_0}
\oplus {\bf 1}_{l},\n\\
&&\Gamma_5 = {\bf 1}_{k} \oplus U_{l_1} \otimes {\bf 1}_{l_0}, \nonumber \\
&&\Gamma_6 = {\bf 1}_{k} \oplus V_{l_1} \otimes {\bf 1}_{l_0}. 
\eea

One can directly check
that the solution (\r{asection}) is compatible with 
the boundary condition (\r{tbcas}) if the matrix $X$ is taken as
\[X=Q\bn\]
where $Q$ and the integer matrix $\bn$ are defined as
\bea
\l{Q}
&&Q^{-1}={\bf 1}_n-{\wt P}\Theta,\\
&&\frac{{\bn^\mu}_\nu}{n}
=\frac{{{N^{(k)}}^\mu}_\nu}{k}{\bf I}_k+
\frac{{{N^{(l)}}^\mu}_\nu}{l}{\bf I}_l,
\eea
and
\bea
\l{N}
&&{{N^{(k)}}^\mu}_\nu=(-m_1 s_2^\mu,\, q_1 s_1^\mu,\, -m_2 s_4^\mu,\,
q_2 s_3^\mu) 
\quad {\rm mod}\quad k, \n\\
&&{{N^{(l)}}^\mu}_\nu=(-m_3 s_6^\mu,\, l_0 s_5^\mu,\,
l\delta^\mu_3,\,l\delta^\mu_4) 
\quad {\rm mod}\quad l.\n
\eea

Let $\cf={\bf 1}_q \otimes {\wt \cf}$. Using Eqs. (\r{F}), (\r{P}), and (\r{M}),
the following identity can be derived 
\bea
\l{Q^2}
Q^2&=&{\bf 1}_n+2\pi  {\wt \cf} \Theta = ({\bf 1}_n-\bm \Theta/n)^{-1},\n\\
&=&{Q^{(k)}}^2{\bf I}_k+{Q^{(l)}}^2{\bf I}_l,
\eea
where
\bea
\l{Qkl}
{Q^{(k)}}^2&=&(1-m^{(k)}\Theta/k)^{-1},\n\\
{Q^{(l)}}^2&=&(1-m^{(l)}\Theta/l)^{-1}.\n
\eea
Using the identity, the constant curvature (\r{F}) can be rewritten as
\be
\l{Fm}
{\wt \cf}=\frac{1}{2\pi}(n{\bf 1}_n-\bm \Theta)^{-1}\bm=
\frac{1}{2\pi}\bm(n{\bf 1}_n-\Theta \bm)^{-1}.
\ee
Then, using the relation \ct{brmozu}
\[\int_{\4t} d^4 x {\rm Tr}\Phi(x)=(2\pi)^4 
\Bigl(k |{\rm det} Q^{(k)}|^{-1}{\rm Tr}_q\Phi_{0000}^{(k)}+ 
l |{\rm det} Q^{(l)}|^{-1}
{\rm Tr}_q\Phi_{0000}^{(l)} \Bigr),\]
where $\Phi_{0000}^{(k)}$ and $\Phi_{0000}^{(l)}$ are the zero modes of 
the expansion (\r{asection}), 
one can check that, as it should be, the 0-brane charge C in (\r{C}) 
is equal to
\be
C=\frac{1}{8\pi^2}\int_{\4t} d^4 x {\rm Tr}\cf \wedge \cf.
\ee 

Now let us calculate the commutation relations staisfied by $Z_\mu$'s,
which are generators of the algebra of functions on a new torus,
denoted by $\4t_{\Theta'}$.
From the explicit form (\r{Z}), the commutation relation of the
generators $Z_\mu$'s can be found as
\be
\l{ZZ}
Z_\mu Z_\nu = e^{2\pi i \Theta'_{\mu\nu}}Z_\nu Z_\mu,
\ee
where
\be
\l{theta'} 
\Theta'=n^{-2}\bn^T Q^T \Theta Q \bn-n^{-1} \bl,
\ee
and the integer matrix $\bl$ is defined by
\bea
\l{L}
&&\frac{\bl_{\mu\nu}}{n}
=\frac{L^{(k)}_{\mu\nu}}{k}{\bf I}_k+
\frac{L^{(l)}_{\mu\nu}}{l}{\bf I}_l,\\
&&L_{\mu\nu}^{(k)}=q_1(s_1^\mu s_2^\nu -s_1^\nu s_2^\mu)
+q_2(s_3^\mu s_4^\nu -s_3^\nu s_4^\mu) \quad {\rm mod} \quad k,\n\\
&&L_{\mu\nu}^{(l)}=l_0(s_5^\mu s_6^\nu -s_5^\nu s_6^\mu) 
\quad {\rm mod} \quad l.\n
\eea
The deformation parameters $\Theta'_{\mu\nu}$ 
on $\4t_{\Theta'}$ given by (\r{theta'}) can be decomposed 
into $U(k)$ part and $U(l)$ part:
\be
\l{newtheta}
\Theta'_{\mu\nu}=\Theta_{\mu\nu}^{'(k)}{\bf I}_k +
\Theta_{\mu\nu}^{'(l)} {\bf I}_l.
\ee  
Here, ${\Theta'}^{(\iota)}\; (\iota=k\,{\rm or}\,l)$ can be rewritten 
as a fractional transformation \ct{brmozu} 
\be
\l{fractr}
{\Theta'}^{(\iota)}=\Lambda_0^{(\iota)}(\Theta)\equiv (A_\iota\Theta+ B_\iota)
(C_\iota\Theta+D_\iota)^{-1},
\ee
where 
\be
\l{lambda}
\Lambda_0^{(\iota)}= \pmatrix{ 
A_\iota & B_\iota \cr
C_\iota & D_\iota }
\ee 
and the four dimensional matrices are defined by
\be
\l{abcd}
A_\iota=n^{-1}_\iota (N^T_\iota + L_\iota N^{-1}_\iota m_{0\,\iota}),
\;\; B_\iota=-L_\iota N^{-1}_\iota,\;\;C_\iota=-N^{-1}_\iota m_{0\,\iota},\;\;
D_\iota=n_\iota N^{-1}_\iota
\ee
with notation $n_k=k,\;n_l=l$.
One can check that each $\Lambda_0^{(\iota)}$ is an element of 
$SO(4,4|{\bf Z})$, which is a T-duality group of the type II string
theory compactified on $\4t$;
\bea
&&\Lambda_0^{(\iota)\,T} J \Lambda_0^{(\iota)}=J,\n \\
&&J = \pmatrix{ 
0 & {\bf 1}_4 \cr
{\bf 1}_4 & 0 }.
\eea

For (4,\,2,\,2) or (4,\,2,\,2,\,0) backgrounds where the magnetic fluxes take the
form of diagonal matrices breaking the gauge group to $U(k)\times U(l)$,
Eq.(\r{newtheta}) implies that the moduli space for the D-brane bound
states is described by two noncommutative parameters 
$\Theta^{'(k)}$ and $\Theta^{'(l)}$.
Thus we expect that it takes the form 
$(\4t_{\Theta^{'(k)}})^p/S_p \times (\4t_{\Theta^{'(l)}})^q/S_q$ 
with $p$ and $q$ determined by ranks and fluxes \ct{dm,hm}.

\section{$SO(4,4|\bz)$ Duality and Morita Equivalence}

In this section we analyze the bound states with nonzero $D0$-brane
charge, $C \neq 0$, corresponding to the (4,\,2,\,2,\,0) system.
For the given fluxes $m_0$ in (\r{Mx}), we take the integral matrices 
$L^{(k)}$ and $L^{(l)}$ to be as close to the inverses of 
$m_0^{(k)}$ and $m_0^{(l)}$ as possible, respectively:  
\bea
\ba{ll}
\l{Lx}
L^{(k)} = \pmatrix{ 
0 & -q_1b_1 & 0 & 0 \cr
q_1b_1  & 0 & 0 & 0 \cr
0 & 0 & 0 &  -q_2b_2\cr
0 & 0 & q_2b_2 & 0 },\quad
&L^{(l)} = \pmatrix{ 
0 & -l_0b_3 & 0 & 0 \cr
l_0b_3  & 0 & 0 & 0 \cr
0 & 0 & 0 & 0  \cr
0 & 0 &  0 & 0 },
\ea
\eea
where $b_1,\,b_2$, and $b_3$ are integers
such that $a_1 k-b_1 m_1=q_1,\;a_2 k-b_2 m_2=q_2$, 
and $a_3 l-b_3 m_3=l_0$, respectively.
Here, we define ${\tilde m}_i=m_i/q_i$ and 
${\tilde m}_3=m_3/l_0$, so that $a_i k_i-b_i {\tilde m}_i=1$
and $a_3 l_1-b_3 {\tilde m}_3=1$.
Then the set of integers $s_\alpha^\mu$ in (\r{L})
can be chosen to satisfy (\r{Lx})
\bea
&& s_1^\mu=(0,\,1,\,0,\,0),\quad s_2^\mu=(b_1,\,0,\,0,\,0),\n\\
&& s_3^\mu=(0,\,0,\,0,\,1),\quad s_4^\mu=(0,\,0,\,b_2,\,0),\n\\
&& s_5^\mu=(0,\,1,\,0,\,0),\quad s_6^\mu=(b_3,\,0,\,0,\,0).
\eea
Also, for the above given set, 
the matrices $N^{(k)}$ and $N^{(l)}$ are
given by
\bea
\l{Nx}
\ba{ll}
N^{(k)} = \pmatrix{ 
q_1 & 0 & 0 & 0 \cr
0  & q_1  & 0 & 0 \cr
0 & 0 & q_2 & 0 \cr
0 & 0 & 0  & q_2 },\quad
&N^{(l)} =l_0 \pmatrix{ 
1  & 0 & 0 & 0 \cr
0  & 1  & 0 & 0\cr
0 & 0 & l_1 & 0 \cr
0 & 0 & 0  & l_1 }.
\ea
\eea

 From (\r{abcd}), the $SO(4,4|\bz)$ transformations $\Lambda_0^{(\iota)}$ in
(\r{lambda}) can be found as 
\bea
\l{Lambda0}
&&\Lambda_0^{(k)} = \pmatrix{ 
a_1{\bf 1}_2   & 0             &  b_1 \ve     & 0             \cr
0              & a_2{\bf 1}_2  & 0            & b_2 \ve       \cr
-{\t m}_1 \ve  & 0             & k_1{\bf 1}_2 & 0             \cr
0              & -{\t m}_2 \ve & 0            & k_2{\bf 1}_2  },\\
\l{Lambda0l}
&&\Lambda_0^{(l)} = \pmatrix{ 
a_3{\bf 1}_2   & 0          &  b_3 \ve     & 0          \cr
0              & {\bf 1}_2  & 0            & 0          \cr
-{\t m}_3 \ve  & 0          & l_1{\bf 1}_2 & 0          \cr
0              & 0          & 0            & {\bf 1}_2  },
\eea
where ${\bf 1}_2$ and $\ve$ are $2 \times 2$ identity and
antisymmetric $(\ve^{12}=- \ve^{21}=1)$ matrices, respectively. 
Since the general solution for an aritrary matrix $m$ 
in (\r{Msl4}) is obtained by $SL(4,\bz)$ transformation $R$, 
the corresponding $SO(4,4|\bz)$ transformations $\Lambda_\iota$
can be given by the set $(Rm_0R^T,\,R N,\,L)$ \ct{brmozu}.
With (\r{lambda}), the $SO(4,4|\bz)$ transformation $\Lambda_\iota$ 
 can be found as 
\be
\l{Lambda}
\Lambda_\iota=\Lambda_0^{(\iota)}\pmatrix{ 
R^T & 0 \cr
0 & R^{-1}}.
\ee

Under the $SO(4,4|\bz)$ transformation 
 (\r{Lambda0}) or (\r{Lambda0l}),
the rank, 6 fluxes, and instanton (eight components altogether) make 
a fundamental multiplet of the Weyl spinor representation of
$SO(4,4|\bz)$ and this multiplet is mapped to Morita equivalent tori
by the action of $SO(4,4|\bz)$ \ct{riesch,sch,brmozu,hove}. 
For convenience, the explicit construction will
be performed only for the $SO(4,4|\bz)$ matrix (\r{Lambda0}) 
since, for the matrix (\r{Lambda0l}), it is essentially similar, 
and so we will drop the index $(\iota)$ from here.

Since the vector and spinor representations of $SO(4,4|\bz)$ are
related by 
\be
\l{8s8v}
S^{-1} \c_i S= {\Lambda_i}^j \c_j, \qquad i,\,j=1, \cdots, 8,
\ee
where the gamma matrices satisfy
\be
\{\c_i,\,\c_j\}=2 J_{ij}, 
\ee
the spinor representation $S(\Lambda)$ corresponding to the
transformation $\Lambda=\Lambda_0\Lambda(R)$ in (\r{Lambda}) is a
product of $S(\Lambda_0)$ corresponding to $\Lambda_0$ and 
$S(R)$ corresponding to $\Lambda(R)$
\be
\l{S}
S(\Lambda)=S(\Lambda_0)S(R).
\ee
On $\4t$, the rank $k$, 6 fluxes $m_{\mu\nu}$, and $U(k)$ instanton number,
$C=Pf(m_{\mu\nu})/k$, make a fundamental multiplet of 
the Weyl spinor representation of $SO(4,4|\bz)$.
We write such an 8-dimensional spinor $\psi$ as
\be
\l{8s} 
\psi=k|0>+\frac{1}{2}m^{\mu\nu} a_\mu^\dagger a_\nu^\dagger |0>
+\frac{C}{4!}\e^{\mu\nu\rho\sigma} a_\mu^\dagger a_\nu^\dagger 
a_\rho^\dagger a_\sigma^\dagger |0>,
\ee
with the fermionic Fock basis defined in the Appendix.
Explicitly we take the spinor basis $\psi_\a\, (\a= 1,\cdots, 8)$ as follows 
\be
\l{8spinor}
\psi_\a= (k, m_{34}, m_{42}, m_{23}, m_{12}, m_{13}, m_{14}, C).
\ee
Using the result in the Appendix, $S(R)$ acts on this spinor as 
\be
\l{sl4}
\psi_0=S(R)\psi=(k,m_2, 0, 0, m_1 , 0, 0, {\wt C}),
\ee
where ${\wt C}=m_1 m_2/k$.
Note that the instanton number ${\wt C}={\t m}_1 {\t m}_2k/k_1k_2$ is
integral since $k_1k_2|k$ \ct{baal}.
Now one can check that, using the result in the Appendix,
$S(\Lambda)$ acts on this spinor as 
\bea
\l{8spn0}
\psi'&=&S(\Lambda_0)S(R)\psi=S(\Lambda_0)\psi_0,\n\\
&=&(k_0,0, 0, 0, 0, 0, 0, 0).
\eea

Since the transformation $S(\Lambda)$ is an isomorphism between Fock
spaces described by quantum number $\psi$, (\r{8spn0}) implies that
the quantum tori with quantum number $\psi$ is (Morita) equivalent to
that of $\psi'$. Similarly, the quantum tori described by the matrix
(\r{Lambda0l}) will be mapped to Morita-equivalent tori 
with quantum number $(l_0, 0, 0, 0, 0, 0, 0, 0)$.
Thus it implies that the moduli space of (4,\,2,\,2,\,0) system as well
as (4,\,2,\,2) system in $U(N)$ super Yang-Mills theory 
can be mapped to $D0$-brane moduli
space and so it takes the form $(\4t_{\Theta^{'(k)}})^{qk_0}/S_ {qk_0}
\times (\4t_{\Theta^{'(l)}})^{ql_0}/S_{ql_0}$. 
This prediction is also consistent with the fact that the moduli space
for the reducible connections takes the form of a product of smaller
moduli spaces \ct{hm}.
For a direct generalization, one can consider a generic constant background 
which breaks $U(N)$ to $\prod_a U(k_a)$. 
Then, we expect that the moduli space of $D$-brane
bound states in Type II string theory takes the form 
$\prod_a (\4t_{\Theta^{'(a)}})^{q_a}/S_{q_a}$.

\section{Conclusion and Comments} 
We studied the modules of D-brane bound states on noncommutative
$\4t$ with non-Abelian constant backgrounds 
and examined the Morita equivalence between them.
We found that the quantum tori with various D-brane charges 
is (Morita) equivalent to that of D0-branes.
For a generic constant background which breaks $U(N)$ to $\prod_a U(k_a)$,
it was shown that the moduli space of D-brane
bound states in Type II string theory takes the form 
$\prod_a (\4t_{\Theta^{'(a)}})^{q_a}/S_{q_a}$.

The construction in this paper has only involved constant 
D-brane backgrounds. 
The noncommutative instantons on $\4t$ 
may share some
properties with noncommutative instantons on ${\bf R}^4$ \ct{nesc}
such as the resolution of small instanton singularity. 
Unfortunately the explicit construction of full instanton modules
seems very hard, not due to the noncommutativeness of the
geometry, but rather due to the non-Abelian properties of instanton connections.
It would be very nice to give a construction also for these
non-Abelian instantons since it was claimed in \ct{20} that the
moduli space of the twisted little string theories of $k\; NS5$-branes
at $A_{q-1}$ singularity \ct{KI}, 
compactified on ${\bf T}^3$ is equal to the
moduli space of $k\; U(q)$ instantons on a noncommuative $\4t$. 

Some interesting problems remain.
The present construction may be generalized to the noncommutative
${\bf K3}$ and instanton solutions on it. 
The instanton configurations
on noncommutative $\4t$ or ${\bf K3}$ should be relevant to the microscopic 
structures of D1-D5 black holes with $B_{NSNS}$ field background,
since the counting of microscopic BPS bound states can be related to
the number of massless fields parameterizing the moduli space of the
bound states \ct{sv}.
It is also interesting since the type IIB string theory on $AdS_3
\times {\bf S}^3 \times X$ with nonzero NS-NS B field along $X$, 
where $X$ is ${\bf K3}$ or $\4t$, corresponds to the conformal 
sigma-model whose target space is the moduli space of instantons 
on the noncommutative $X$ \ct{ads}.

Another interesting problem is the deformation quantization of 
Matrix theory on noncommutative $\4t$ \ct{sw}. 
Although the algebra of functions on $\4t$ is deformd by so-called $*$
product, the fuctions can be Fourier expanded in the usual way.
In that case, $*$ product between Fourier expanded functions will be
relatively simple. We hope to address these problems soon.

\section{Appendix}

To construct the spinor representation $S(\Lambda)$, we introduce
fermionic operators $a_\mu^\dagger=\gamma_\mu/\sqrt 2$ and
$a_\mu=\gamma_{4+\mu}/\sqrt 2$ satisfying anti-commutation relations
\be
\{a_\mu,\, a_\nu^\dagger\}=\d_{\mu\nu}, \quad 
\{a_\mu^\dagger,\, a_\nu^\dagger\}=\{a_\mu,\, a_\nu\}=0, 
\quad \mu,\,\nu=1, \cdots,4.
\ee 
Since the $SL(4,\bz)$ transformation does not affect the rank and
the instanton number and the $SL(4,\bz)$ is isomorphic to
$SO(3,3|\bz)$, we expect, in the spinor basis (\r{8s}),
that the spinor representation $S(R)$
corresponding to $\Lambda(R)$ in (\r{Lambda}) has the following form
\be
\l{SR}
S(R) = \pmatrix{ 
1& 0 & 0 \cr
0  & SO(3,3|\bz) & 0 \cr
0 & 0 & 1}.
\ee
Indeed, according to \ct{sch}, the operator ${\bf \Lambda}(R)$
corresponding to $\Lambda(R)$ is given by 
\be
{\bf \Lambda}(R)={\rm exp}(-a_\mu \lambda^{\mu\nu} a_\nu^\dagger),
\qquad (R)_{\mu\nu}={\rm exp}(\lambda_{\mu\nu}),
\ee
and then the spinor representation $S_{\a\b}(R)$ is defined as
\be
{\bf \Lambda}(R)|\b>=\sum_{\a=1}^8 |\a>S_{\a\b}(R). 
\ee
Obviously, acting on the rank $(\b=1)$ and the instanton $(\b=8)$
basis, $S_{\a1}(R)=S_{1\a}(R)=\d_{\a1}$ and $S_{\a8}(R)=S_{8\a}(R)=\d_{\a8}$.
After a little algebra, we can find the $6 \times 6$ matrix in
(\r{SR}) denoted as $H(R)=H_3 H_2 H_1\in SO(3,3|\bz)$
\bea
&&H_1=\pmatrix{ 
C_{12}^T & 0 \cr
0 & C_{12}^{-1}}, \quad
H_2=\pmatrix{ 
{\bf 1}_3 & 0 \cr
\ca & {\bf 1}_3},\quad
H_3=\pmatrix{ 
{\bf 1}_3 & \cb \cr
0 & {\bf 1}_3},\n\\
&&\ca = \pmatrix{ 
0      & -R_{14} & R_{13}\cr
R_{14} &  0      & -R_{11} \cr
-R_{13}&  R_{11} & 0}, \quad
\cb = \pmatrix{ 
0       & d_{14} & d_{13}\cr
-d_{14} &  0     & 0     \cr
-d_{13} &  0     & 0      },
\eea
where $C_{\mu\nu}$ is a $3 \times 3$ matrix formed by removing
$\mu$-th row and $\nu$-th column from the $4 \times 4$ matrix $R$,
$d_{\mu\nu}={\rm det}(C_{\mu\nu})$, and
we normalized the matrix $C_{12}$ to be $SL(3,\bz)$ by absorbing
determinant factor in the above definition.

Next we will construct the spinor representation $S(\Lambda_0^{(k)})$
corresponding to $\Lambda_0^{(k)}$ in (\r{Lambda0}). Let us make a block-wise
Gauss decomposition of $\Lambda_0^{(k)}$
\bea
\Lambda_0^{(k)}&=&\pmatrix{ 
{\bf 1}_4 & 0 \cr
\cc & {\bf 1}_4}\cdot
\pmatrix{ 
G & 0 \cr
0 & G^{-1} }\cdot
\pmatrix{ 
{\bf 1}_4 & \cd \cr
0 & {\bf 1}_4},\n\\
&=&\Lambda_{\cc}\cdot\Lambda_{G}\cdot\Lambda_{\cd},
\eea
where antisymmetric matrices $\cc,\,\cd$ and a symmetric matrix $G$
are given by
\be
\cc=-\pmatrix{ 
\frac{m_1}{a_1}\ve & 0 \cr
0 & \frac{m_2}{a_2}\ve}, \quad
\cd=\pmatrix{ 
\frac{b_1}{a_1}\ve & 0 \cr
0 & \frac{b_2}{a_2}\ve}, \quad
G=\pmatrix{ 
a_1{\bf 1}_2 & 0 \cr
0 & a_2{\bf 1}_2},
\ee
and $\ve$ is an antisymmetric $2\times 2$ matrix.
Then the corresponding spinor operator ${\bf \Lambda}_0^{(k)}$ will be given by
\be
{\bf \Lambda}_0^{(k)}
={\rm exp}(\frac{1}{2}\cc^{\mu\nu}a_\mu^\dagger a_\nu^\dagger)\cdot
{\rm exp}(-h^{\mu\nu} a_\mu^\dagger a_\nu)\cdot
{\rm exp}(\frac{1}{2}\cd^{\mu\nu}a_\mu a_\nu), 
\ee
where $(G)_{\mu\nu}={\rm exp}(h_{\mu\nu})$.
Thus the representation $S(\Lambda_0^{(k)})$ can be obtained by a product of each
spinor representation,
\be
S(\Lambda_0^{(k)})=S(\Lambda_{\cc})\cdot S(\Lambda_{G})\cdot
S(\Lambda_{\cd}),
\ee
where
\be
\l{SLambda0}
S(\Lambda_0^{(k)}) = \pmatrix{ 
a_1a_2          & -a_1b_2     &    0       & 0    
& -a_2b_1       & 0           &    0       & b_1b_2 \cr
-a_1{\t m}_2    & a_1k_2      &    0       & 0   
&b_1{\t m}_2    & 0           &    0       & -b_1k_2  \cr
0               & 0           &    1       & 0   
&   0           & 0           &    0       & 0 \cr
0               & 0           &    0       & 1 
&   0           & 0           &    0       &  0  \cr
-a_2{\t m}_1    & b_2{\t m}_1 &    0       &  0  
&  a_2 k_1      & 0           &    0       & -b_2 k_1   \cr 
0               & 0           &    0       &  0  
&   0           & 1           &    0       & 0   \cr
0               & 0           &    0       & 0 
&   0           & 0           &    1       & 0   \cr
{\t m}_1{\t m}_2&-{\t m}_1k_2 &    0       &  0  
& -{\t m}_2 k_1 & 0           &    0       & k_1 k_2 }.
\ee
Similarly,
\be
S(\Lambda_0^{(l)}) = \pmatrix{ 
a_3 & 0  & 0    & 0   & -b_3 & 0   & 0     & 0    \cr
0   & a_3& 0    & 0   &  0   & 0   & 0     & -b_3 \cr
0   & 0  & 1    & 0   &  0   & 0   & 0     & 0    \cr
0   & 0  & 0    & 1   &  0   & 0   & 0     & 0    \cr
-{\t m}_3& 0  & 0    & 0   &  l_1   & 0   & 0     & 0    \cr 
0   & 0  & 0    & 0   &  0   & 1   & 0     & 0    \cr
0   & 0  & 0    & 0   &  0   & 0   & 1     & 0    \cr
0   &-{\t m}_3& 0    & 0   &  0   & 0   & 0     & l_1  }.
\ee
Here we used the definition (\r{8s8v}) in order to drop
the global factors such as $1/a_1a_2$ and $1/a_3$.

\vspace{.5cm}
\noindent
{\large\bf Acknowledgments}\\[.2cm]
We would like to thank APCTP for its kind hospitality, where part of
this work was done. E. K., H. K. and C.-Y. L. would like to thank
KIAS for its kind hospitality, where part of this work was done 
while they were visiting. 
B.-H. L. and H. S. Y. were supported by the Ministry of Education, BK21
Project No. D-0055 and by grant No. 1999-2-112-001-5 from the
Interdisciplinary Research Program of the KOSEF. 
E. K., H. K. and C.-Y. L. were supported 
in part by KRF, Interdisciplinary Research 
Project 1998-D00001.
C.-Y. L. was also supported in part by
BSRI Program 1998-015-D00073 and NSF PHY-9511632.

\newcommand{\J}[4]{{\sl #1} {\bf #2} (#3) #4}
\newcommand{\andJ}[3]{{\bf #1} (#2) #3}
\newcommand{\AP}{Ann.\ Phys.\ (N.Y.)}
\newcommand{\MPL}{Mod.\ Phys.\ Lett.}
\newcommand{\NP}{Nucl.\ Phys.}
\newcommand{\PL}{Phys.\ Lett.}
\newcommand{\PR}{Phys.\ Rev.}
\newcommand{\PRL}{Phys.\ Rev.\ Lett.}
\newcommand{\CMP}{Comm.\ Math.\ Phys.}
\newcommand{\JMP}{J.\ Math.\ Phys.}
\newcommand{\JHEP}{J.\ High \ Energy \ Phys.}
\newcommand{\PTP}{Prog.\ Theor.\ Phys.}
\newcommand{\ib}{{\it ibid.}}
\newcommand{\hep}[1]{{\tt hep-th/{#1}}}


\end{document}